\documentclass[journal]{IEEEtran}

\ifCLASSINFOpdf
\else
   \usepackage[dvips]{graphicx}
\fi
\usepackage{url}

\usepackage{graphicx}
\usepackage{tikz}
\usetikzlibrary{arrows, decorations.markings}
\usepackage{amsmath}
\usepackage{booktabs}
\usetikzlibrary{spy,positioning,calc}
\usepackage{pgfplots,pgfplotstable}
\pgfplotsset{compat=1.8}

\DeclareMathOperator*{\argmax}{arg\,max}

\newcommand{\ie}{\emph{i.e.}}
\newcommand{\eg}{\emph{e.g.}}

\begin{document}

\title{Video to All-in-focus Image Reconstruction Algorithm for Automated Microscopic Urinalysis\\ }

\author{Chinmay Nema$^*$, Hari Om Aggrawal$^*$, Dipam Goswami, Rajiv Gupta, Vinti Agarwal
\thanks{Date:}
\thanks{Chinmay Nema, Dipam Goswami, and Vinti Agarwal are with Birla Institute of Technology and Science Pilani, India. Hari Om Aggrawal is an independent researcher, India; previously with the Institute of Mathematics and Image Computing, University of Lübeck, Germany. Rajiv Gupta is with Gurudwara Singh Sabha Charitable Dispensary, Rajpura, India.}
\thanks{Corresponding authors: C. Nema (e-mail: chinmaynema5230@gmail.com) and H. O. Aggrawal (email: hariom85@gmail.com). \\ *equal contribution\protect}
}

\markboth{Journal of \LaTeX\ Class Files, Vol. 14, No. 8, August 2015}
{Shell \MakeLowercase{\textit{et al.}}: Bare Demo of IEEEtran.cls for IEEE Journals}
\maketitle

\begin{abstract}
Microscopic urinalysis is a routine diagnostic test at hospitals. Recent studies have demonstrated the effectiveness of deep learning methods to automate microscopic urinalysis. These methods rely on high-quality images of the urine samples in which each cell is clearly identifiable. However, in practice, the urine sample on a glass slide has a multi-layer structure; hence, all the cells are not clearly visible within the depth of field of a lens focused at a particular focal plane.
It demands acquiring multiple images at different focal planes to correctly identify each cell in a given urine sample, which is  a time-consuming task.

In this paper, we propose to simplify the task by recording a video, in place of acquiring multiple images, while gradually changing the focus of the lens manually by hand. A typical length of the video is from 2 to 14 seconds. We reconstruct an all-in-focus image from the recorded video frames and apply a deep learning model to detect and classify urine sediments. As a proof of concept, we conduct experiments on 14 videos acquired by a trained lab technician in a usual diagnostic lab environment and show the effectiveness of the proposed automated urinalysis pipeline with our novel reconstruction algorithm.

\end{abstract}


\IEEEpeerreviewmaketitle

\section{Introduction}

	
Urine is a rich source of information. It is an asset for doctors to provide early signs of urinary tract infections (UTI) \cite{Jeff2005}, renal failure \cite{Susan2005} etc. Every year, 3\% of people die due to renal failure \cite{Jha2017}, and 19\% of pregnant women suffer from acute symptoms of UTI in India \cite{Band2005}. Untreated UTI cases can lead to kidney infections, premature rupture of membrane, and low birth weight baby \cite{Ky2007}.

Microscopic urine sediment analysis is a gold-standard method for urinalysis \cite{Delanghe2014}. 
It is a rapid, cost-effective, and reliable diagnostic test if performed timely and carefully.
Analyzing urine sediments such as pus cells, red blood cells (RBCs), epithelial, cast, crystals in a urine image requires special expertise that is not always available in pathology labs \cite{Oyaert2019}. Hence, many manufactures provide automated urine microscopy analyzers that classify and count cells without any human intervention \cite{Oyaert2019, Aziz2018}.
Numerous comparative studies have shown the effectiveness of analyzers over manual methods \cite{Sharda2014, nce2016}. 

The commercial urine analyzers generally come as a stand-alone solution with an in-built microscope, an autofocus optical system, camera, and processing unit. These systems are expensive and suited for large-sized labs for testing in large numbers. In this work, we automate urinalysis without an autofocus system mainly for small-sized labs and to handle the shortage of experienced staff in villages.

The autofocus system is an essential component of urine analyzers. In practice, the urine sample on a glass slide has a multi-layer structure; due to that, all the cells are not clearly visible within the depth of field of a lens focused at a particular focal plane. Hence, images are acquired at multiple focal planes to identify each cell in a given urine sample. Fine focus adjustments are necessary to obtain high-quality images which is a time-consuming task without an autofocus system. 

\begin{figure}[t]
    \centering
    \includegraphics[width=\columnwidth]{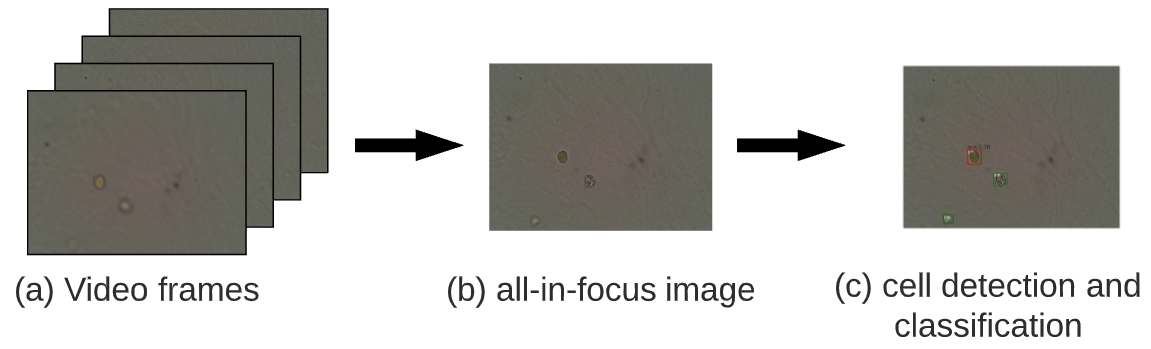}
    \caption{The proposed video-based automated microscopic urinalysis pipeline. The contrast of urine microscopic images are generally very low; they are best visible only in the digital documents.}
    \label{fig:pipeline}
\end{figure}

In this paper, we propose to simplify the focus adjustment task by recording a video while gradually changing the focus of the lens manually by hand. We then use our novel all-in-focus reconstruction algorithm to obtain a single composite image containing all the cells in the sample, and finally use a neural network to identify and classify many types of cells in the sample; see our urinalysis pipeline in Fig.~\ref{fig:pipeline}. 

Our video acquisition strategy requires only a smartphone or a camera adapter that connects with the microscope eyepiece. It is a cost-effective alternative to an expensive autofocus system. The proposed strategy is user-friendly and easy to adopt. It does not require any special training for lab technicians. 

In this study, the lab technician recorded videos of 14 urine samples and achieved up to 95\% positive predictive value, 70\% true positive rate through our microscopic urinalysis pipeline.

The structure of the paper is as follows: Sec.~\ref{Sec:relatedwork} covers the related work to all-in-focus image reconstruction and deep learning methods for urinalysis. Sec.~\ref{Sec:algo} describes our patch-based all-in-focus image reconstruction algorithm. Sec.~\ref{Sec:results} reports the performance of our pipeline for urinalysis. 
In Sec.~\ref{Sec:conclusion}, we conclude our findings.

\newcommand{\drawlinenboxes}[5]{
\begin{scope}[#1]
\draw[help lines,xstep=.125,ystep=.167] (0,0) grid (1,1);
\coordinate (#2) at (0.45,0.3);
\coordinate (#3) at (0.85,0.8);
\spy[green,size=0.7cm] on (#2) in node at (#3);
\coordinate (#4) at (0.3375,0.45);
\coordinate (#5) at (0.15,0.8);
\spy[black,size=0.7cm] on (#4) in node at (#5);
\end{scope}
}

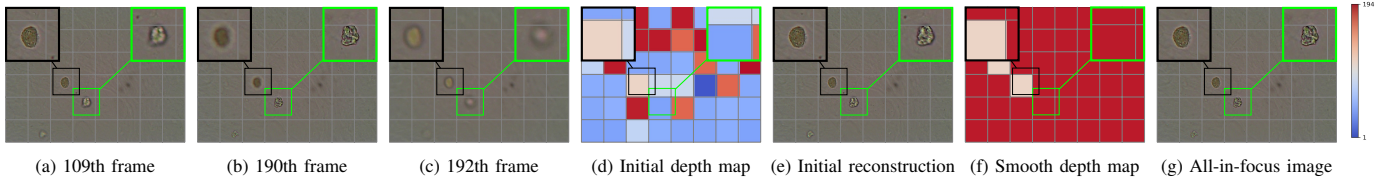
\begin{figure*}[t]
    \centering
    \begin{tikzpicture}[x=0.14\textwidth,font=\scriptsize,spy using outlines={magnification=2, size=2pt}, connect spies]
        \node[anchor=south west,inner sep=0] (a) at (0,0) {\includegraphics[width=0.13\textwidth]{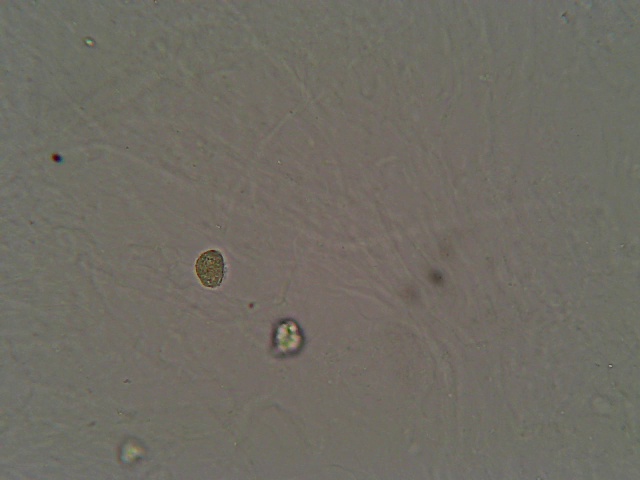}};
        \node[below=0.1 cm of a] {(a) 109th frame};
        \drawlinenboxes{x={(a.south east)},y={(a.north west)}}{p1}{p2}{c1}{c2}
        \node[anchor=south west,inner sep=0] (b) at (1,0)
        {\includegraphics[width=0.13\textwidth]{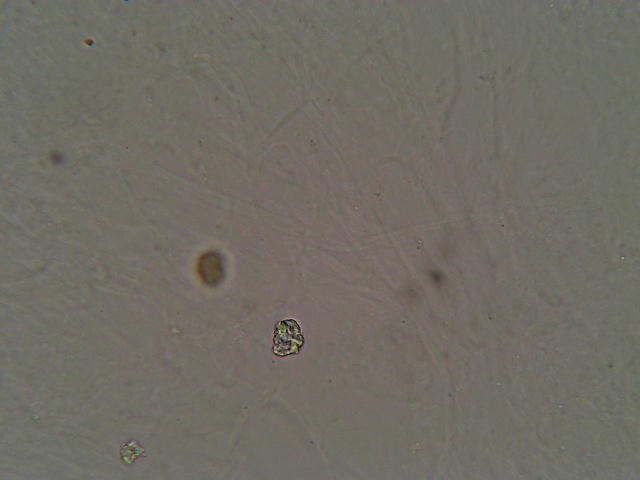}};
        \node[below=0.1 cm of b] {(b) 190th frame};
        \drawlinenboxes{shift={(1,0)},x={(b.south east)},y={(b.north west)}}{p3}{p4}{c3}{c4}
        \node[anchor=south west,inner sep=0] (c) at (2,0)
        {\includegraphics[width=0.13\textwidth]{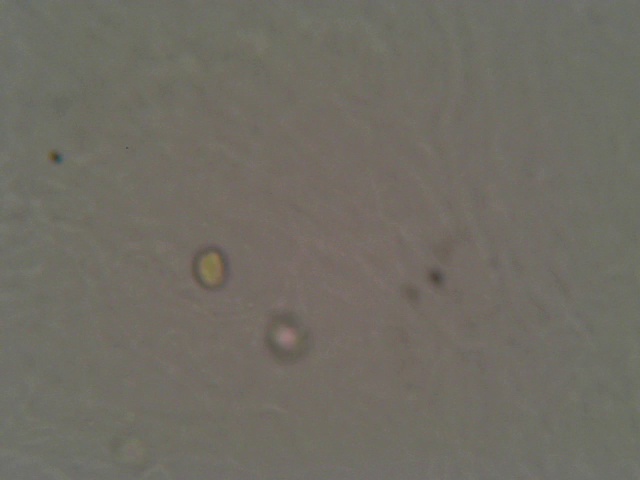}};
        \node[below=0.1 cm of c] {(c) 192th frame};
        \drawlinenboxes{shift={(2,0)},x={(c.south east)},y={(c.north west)}}{p5}{p6}{c5}{c6}
        \node[anchor=south west,inner sep=0] (d) at (3,0)
        {\includegraphics[width=0.13\textwidth]{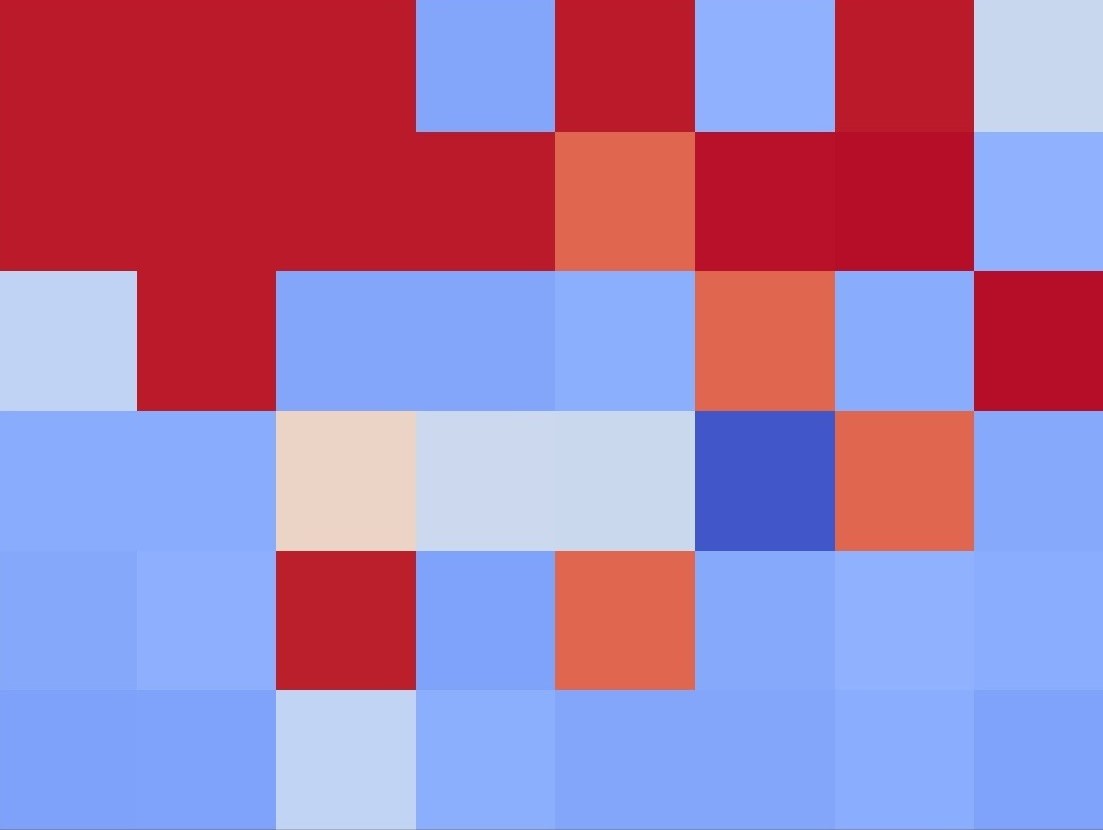}};
        \node[below=0.1 cm of d] {(d) Initial depth map};
        \drawlinenboxes{shift={(3,0)},x={(d.south east)},y={(d.north west)}}{p7}{p8}{c7}{c8}
        \node[anchor=south west,inner sep=0] (e) at (4,0)
        {\includegraphics[width=0.13\textwidth]{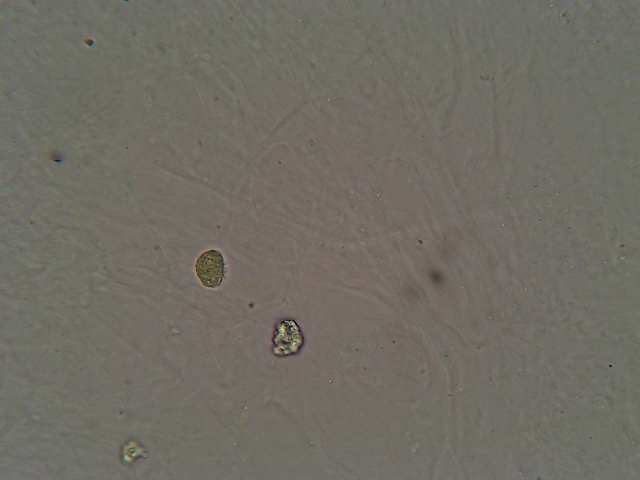}};
        \node[below=0.1 cm of e] {(e) Initial reconstruction};
        \drawlinenboxes{shift={(4,0)},x={(e.south east)},y={(e.north west)}}{p9}{p10}{c9}{c10}
        \node[anchor=south west,inner sep=0] (f) at (5,0)
        {\includegraphics[width=0.13\textwidth]{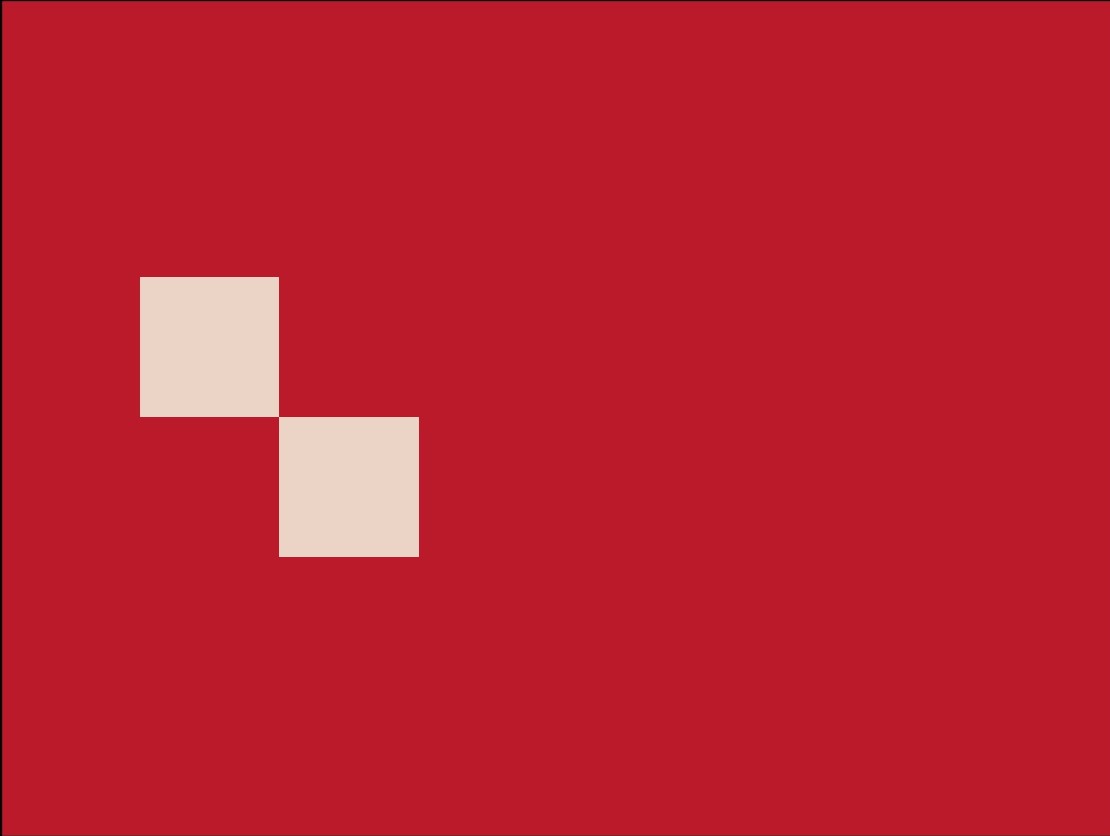}};
        \node[below=0.1 cm of f] {(f) Smooth depth map};
        \drawlinenboxes{shift={(5,0)},x={(f.south east)},y={(f.north west)}}{p11}{p12}{c11}{c12}
        \node[anchor=south west,inner sep=0] (g) at (6,0)
        {\includegraphics[width=0.13\textwidth]{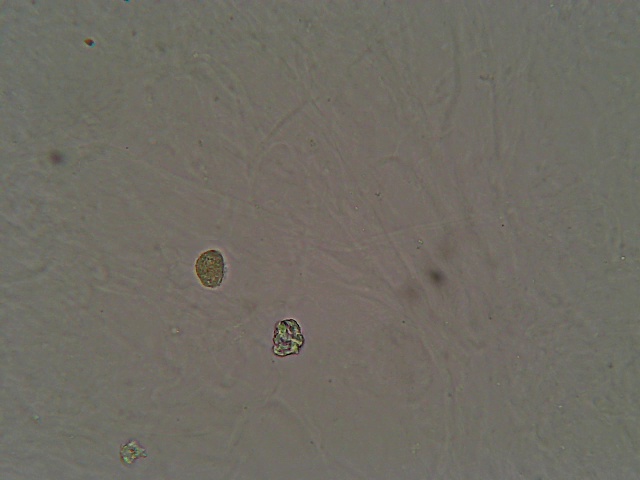}};
        \node[below=0.1 cm of g] {(g) All-in-focus image};
        \drawlinenboxes{shift={(6,0)},x={(g.south east)},y={(g.north west)}}{p13}{p14}{c13}{c14}
        \node[anchor=south west,inner sep=0] (g) at (7,0)
        {\includegraphics[width=0.0185\textwidth]{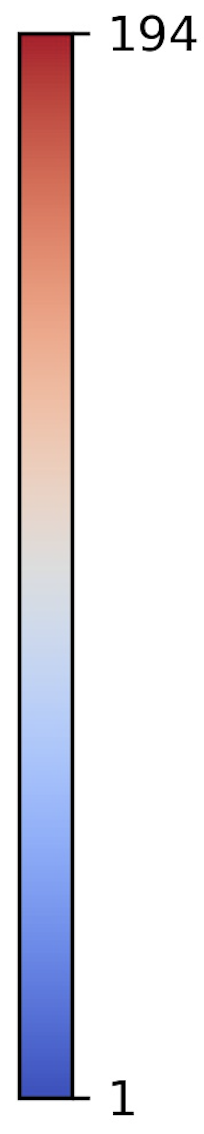}};
    \end{tikzpicture}
    \caption{An example of the steps of our reconstruction algorithm. We first divide each video frame into square patches. Here, out of total 194 frames, we show only three frames 109th (a), 190th (b), and 192nd (c), to illustrate the multi-layer structure of a urine sample. The cell in black box is sharp in the 109th frame, whereas the cell in green box is sharp in the 190th frame. After that, we estimate an initial depth map (d), that is generally noisy which leads to discontinuities in the reconstruction (e). After smoothening of depth map (f), we obtain a sharp all-in-focus image (g) having all cells in focus. 
    Each grid of the depth map indicates the frame index that is used to reconstruct the corresponding patch.}
    \label{fig:recnsteps}
\end{figure*}

\section{Related Work}\label{Sec:relatedwork}

We review the state-of-the-art methods for all-in-focus image reconstruction and report the recent progress in automated microscopic urinalysis with deep learning methods.\\
~\\
\textbf{All-in-focus image reconstruction:}
In this work, we mainly follow the literature of focus stacking algorithms that aim to reconstruct an image from multiple images. The objects in the reconstructed image have a greater depth of field. The existing approaches determine a depth map at each pixel that encodes the position of a maximally focused pixel in the given image sequence. 

The depth map’s accuracy depends on the sharpness measures that tell us the amount of focus at each pixel. It has been noted that 
one measure may not work for all types of images. An extensive survey by \cite{Pertuz2013} compares more than 30 focus measures.
The initial depth map is generally quite noisy, especially in the low-textured regions. Therefore, numerous approaches exist that regularize the depth map by incorporating prior information \cite{Moeller2015,Salokhiddinov2020}.


Focal stacking algorithms are usually designed to cater to specific applications. For example, Focus ALL \cite{Sigdel2016} method is verified on microscopic protein crystallization images. Deep Focus method \cite{Goldsmith2011} reconstruct composite images of surface fractures. The Zerene stacker \cite{zerene} software designed specifically for challenging macro subjects; this software produces extremely unsatisfactory results with our urine sample images. To the best of our knowledge, there have been no studies on focus stacking for urine microscopic images.
\\
\textbf{Deep learning methods for urinalysis:}
Numerous studies \cite{Liang2018, Liang1-2018, Xu2019UrineSD} reported the performance of state-of-the-art object detection models on urine sediment classification and detection. These studies use high-quality images to train neural networks and obtain upto $80\%$ classification accuracy. 





	
\begin{section}{Reconstruction Algorithm}\label{Sec:algo}

%



In this section, we present a simple but effective algorithm to reconstruct an all-in-focus image from video frames. Video is acquired while varying the focus of the lens manually over time. We estimate an all-in-focus image where all cells are in-focus and sharp so that the neural network can identify the features of the cells and classify and detect them accurately. We mainly follow the standard focus stacking steps; see Fig.~\ref{fig:recnsteps} for the steps of our reconstruction scheme. 

In a microscope, the viewing direction from the objective lens to the glass slide is almost perpendicular to the horizontally placed glass slide. Hence, while adjusting the focus of lens, we can assume that the effect of focus is almost similar in a large field of view. Hence, instead of analyzing the sharpness pixel-wise, we calculate it patch-wise that tells us the amount of focus over a large region.

We compute sharpness by the Frobenius norm of Hessian of image intensity for each patch. Hessian outperforms over gradient and Laplacian on urine images based on our experiments. After that, we estimate an initial depth map. Let $(I_n)_{n=1}^{N}$ be a sequence of $N$ two-dimensional video frames. After computing sharpness measure $f(x,y)$ at each patch location $(x,y)$ in each frame, we define a depth map $d(x,y)$ with the frame position where the amount of focus is maximal and reconstruct an all-in-focus image $\hat{I}(x,y)$ given by
\[
    \hat{I}(x,y) = I_{\hat{d}}(x,y) \quad \mbox{where} \quad \hat{d}(x,y) = \argmax_{n} f_n(x,y)
\]
where $1 \leq n \leq N$.
In practice, this simple approach provides a very noisy depth map; \eg~see Fig~\ref{fig:recnsteps}(d). In the background and low-textured regions, the sharpness measures do not have a well-defined peak. 
Moreover, even in the textured regions, the surrounding patches could have maximal focus from very different frames due to minor numerical differences in the sharpness measure. Our video acquisition scheme does not put restrictions on performing the focus tuning cycle more than once. Hence, the depth map in the neighboring patches could have values from different cycles.

For smoothening of depth map in the local neighbourhood, we set $d(i+p,j+q) = d(i,j)$ where $p \in \{-1,0,1\}$ and $q \in \{-1,0,1\}$ if the sharpness measure satisfy one of the following inequalities
\begin{align*}
    &|f_k(i+p,j+q) - f_m(i+p,j+q)|/|f_k(i+p,j+q)| < \tau, \\
    &f_m(i+p,j+q) > f_k(i+p,j+q)
\end{align*}
where $k = d(i+p,j+q)$ and $m = d(i,j)$ and $\tau$ is a hyper parameter. We repeat smoothening steps until the number of unique frame indices in the depth map stops decreasing.

Patch size and $\tau$ are the only hyperparameters in our scheme. Their appropriate selection ensures sufficient sharpness and continuity among neighboring patches. If all of the cells are in focus at the same focal plane, the patch size equal to the video frame size is the best choice. Otherwise, a smaller patch size is preferable to detect localized in-focus regions at different depths of the video. With the smaller patch, we might observe discontinuity in the reconstructed image if the depth map is not locally smooth. For that, we increase threshold $\tau$ to improve smoothness, but it can reduce the sharpness of a few patches in the neighborhood. Discontinuity is mainly noticeable if the reconstructed object is a composite of patches from very different frames due to non-smooth depth map.

Let $w$ and $h$ be the width and height of video, we define patch size $s=\min\{w, h\}/p$ where $p$ is the tuning parameter.

 

\end{section}
\section{Experiments and Results}\label{Sec:results}

In this proof of concept study, we show the effectiveness of our approach on 14 videos collected by lab technician in the normal diagnostic lab conditions. These videos are very short, varying from 2 to 14 secs and easy to record through normal camera attached to a microscope; see hardware and dataset specifications in Sec.~\ref{sec:hspecs} and Sec.~\ref{sec:dataspecs} respectively.

Our automated urinalysis pipeline; see Fig.~\ref{fig:pipeline}, is able to detect 70\% of the cells present at different frames of a video.

\subsubsection{Hardware specifications}\label{sec:hspecs}
We acquired videos from two microscopes but with the same model number CH20iBIMF, Magnus Opto Systems India Pvt. Ltd, connected to a COSLAB 5 megapixel digital camera (model COSUSB5000) through one of the eyepieces of the microscope. Both the camera reduction lens (0.5x) and achromatic objective lens (40x) add to the total magnification of the urine sample.

\subsubsection{Videos and annotations}\label{sec:dataspecs}
We instructed the lab technician to record at least one full cycle of blur, \ie, blur-sharp-blur. But, the recorded videos have 1 to 2 cycles of blur. Moreover, the rate of change of focus varies for each video. Hence, the length of each video is different; see Tab.~\ref{tab:video}. Videos were collected at multiple resolutions based on technician's preference on the image quality to differentiate cells. H.264 video codec was used for compression.

Two authors of this paper annotated the cells and verified them with a doctor. In our case, annotating cells in each video frame is a redundant task. Due to our acquisition strategy, mostly we observe the same cell in multiple frames but with different degree of blur.
Hence, we annotate only few frames that have the sharpest occurrences of cells present in the video. Our reconstruction algorithm assumes that the cells are stationary over the time. Except video 13, the recorded videos follows our stationarity assumption. 

The dataset will be available on Github with ground truth annotations. These annotations are compared with predictions from the neural network to evaluate the performance of our 
pipeline.

\subsubsection{Deep neural network (DNN)}\label{sec:dnn}
In our work, we train RetinaNet \cite{Lin2020}, a popular one-stage object detection model with images, having 3733 urine sediments, available publicly at \cite{UMID}. These images are acquired by adjusting the focus of lens manually. Hence, the dataset contains both sharp and out of focus cells. The performance of the trained model on testing images given in \cite{UMID} is satisfactory. The trained model achieves F1 score 0.86 on RBC, 0.85 on pus, and 0.80 on epithelial cells at confidence level 0.6 on the testing dataset.

Note that, the DNN is trained for images not videos. Hence, we apply DNN on the all-in-focus reconstructed image to obtain the final cell detections and compare them with ground truth annotations. We quantify the performance both visually and quantitatively in terms of precision, recall, and F1 score.

\begin{table}[t]
    \caption{Summary: video properties (length, width, and height), hyperparameters (threshold $\tau$ and patch factor $p$), the number of frames used to reconstruct images after 'initial' estimate of depth map and after smoothing of depth map ('final') out of the 'total' frames in the video, and the ground truth counts of epithelial(ep), pus, and RBC cells in each video.}
	\label{tab:video}
	\setlength{\tabcolsep}{2.5pt}
	\centering
		\begin{filecontents*}{data/all_in_focus_info.csv}
,video_name,video_duration,total_frames,frame_width,frame_height,frame_count_depth_map,frame_count_final_image,number_of_iterations,threshold,patch_size,ep_ground_truth,pus_ground_truth,rbc_ground_truth
0,simplest_ML,13.556000000000001,194,480,640,27,2,288,0.15,6,0,3,0
1,rbc_pus_ML,8.985,75,720,1280,41,2,396,0.15,6,0,1,11
2,ML_86,2.574,57,600,800,17,4,288,0.15,6,3,19,3
3,ML_55,4.4,94,600,800,34,3,288,0.15,6,1,2,0
4,simple_ML,7.0200000000000005,101,480,640,34,2,288,0.15,6,0,2,1
5,ML_121,6.443,54,720,1280,22,4,396,0.15,6,0,0,1
6,ML_118,9.36,78,720,1280,28,4,396,0.15,6,0,2,0
7,ML_117,10.374,24,1944,2592,18,2,288,0.15,6,0,0,1
8,SL_75,3.1199999999999997,69,600,800,21,3,288,0.15,6,4,17,0
9,SL_66,3.4159999999999995,29,720,1280,11,1,192,0.15,4,0,0,4
10,SL_87,2.668,59,600,800,17,5,288,0.15,6,3,8,0
11,SL_12,4.072,90,600,800,19,1,144,0.15,4,3,0,0
12,SL_77,2.324,52,600,800,19,5,288,0.15,6,4,20,0
13,ForPaper_2,13.634,298,600,800,6,2,36,0.05,2,4,4,0
\end{filecontents*}
\pgfplotstabletypeset[
		col sep = comma,
		column type = c,
		columns/video_duration/.style={numeric type, precision=2,zerofill},
		columns/threshold/.style={numeric type, fixed, precision=2,zerofill},
		columns = {[index]0,[index]2,[index]4,[index]5,[index]9,[index]10,[index]3,[index]6,[index]7,[index]11,[index]12,[index]13},
		every head row/.style={%
	        output empty row,
			before row={
			\toprule
			&\multicolumn{3}{c}{Video} & \multicolumn{2}{c}{Parameters} & \multicolumn{3}{c}{Frames} & \multicolumn{3}{c}{Cell counts}\\
			\cmidrule(l){2-4} \cmidrule(l){5-6} \cmidrule(l){7-9} \cmidrule(l){10-12}
			No & Length & Width & Height & $\tau$ & $p$ & Total & Initial & Final & ep & pus & RBC \\
			& (sec.) & $w$ & $h$ & & & & & & & & \\
			},
			after row = {
			\midrule
			}},
		every last row/.style={after row=\bottomrule},
		]{data/all_in_focus_info.csv}		
\end{table}

\subsection*{Overall Performance}

\begin{figure*}
\def\width{0.33\linewidth}
\newcommand{\ig}[1]{\includegraphics[width=\width]{#1}}
    \centering
    \begin{tikzpicture}[x=0.33\linewidth,y=0.17\linewidth,every node/.style={inner sep=2pt,outer sep=0pt}, scale=1]
		\node[label={above: (a) ML Video},label={[anchor=south, rotate=90]left:}] at (0,0) {\ig{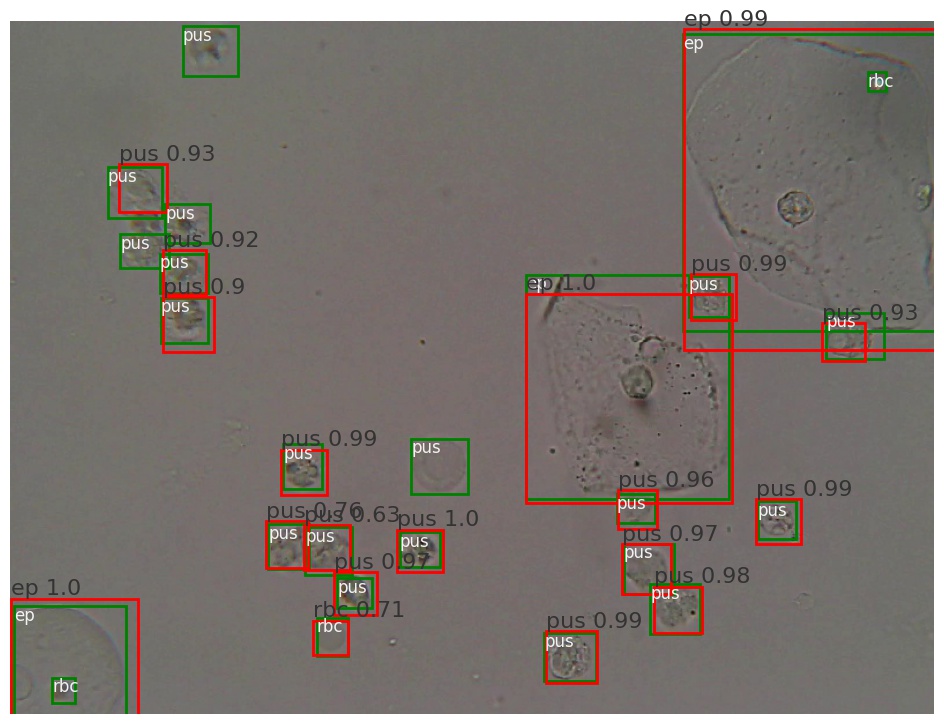}};
		\node[label={above: (a) SL Video},label={[anchor=south, rotate=90]left:}] at (1,0) {\ig{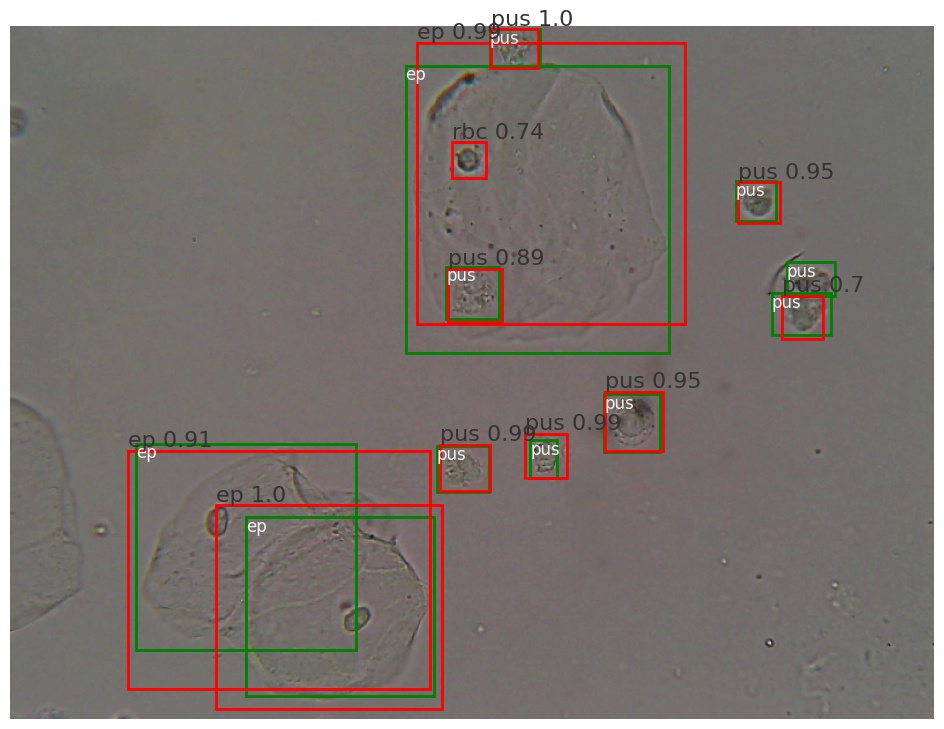}};
		\node[label={above: (a) MV Video},label={[anchor=south, rotate=90]left:}] at (2,0) {\ig{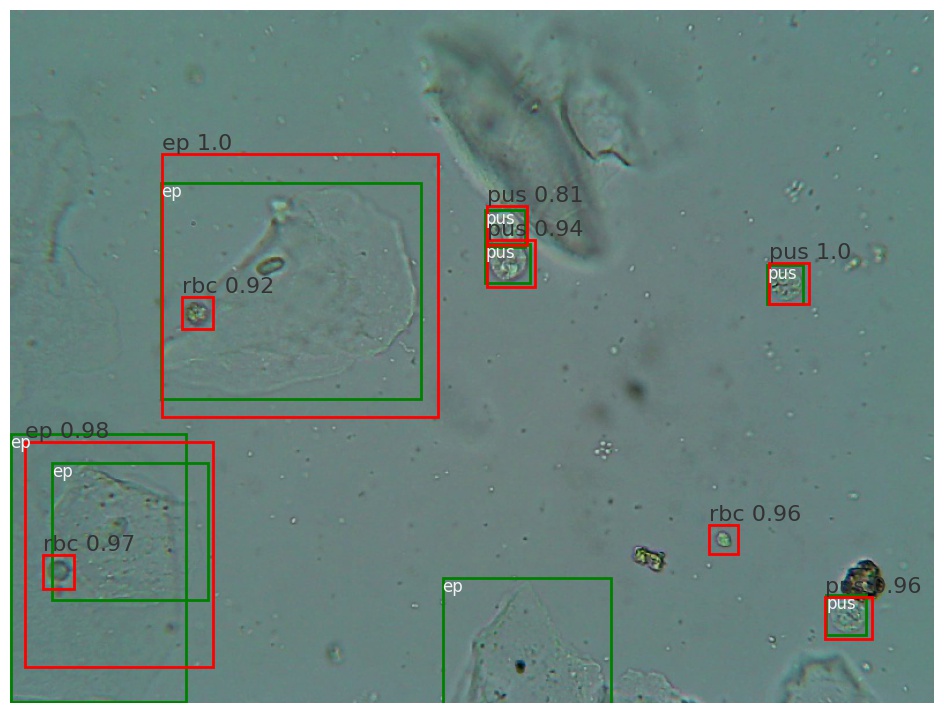}};
	\end{tikzpicture}
\def\width{0.15\linewidth}
    \begin{tikzpicture}[x=0.16\linewidth,y=0.14\linewidth,every node/.style={inner sep=2pt,outer sep=0pt}, scale=1, font=\footnotesize]
		\node[label={above: $0^{th}$ frame},label={[anchor=south, rotate=90]left:ML Video }] at (0,0) {\ig{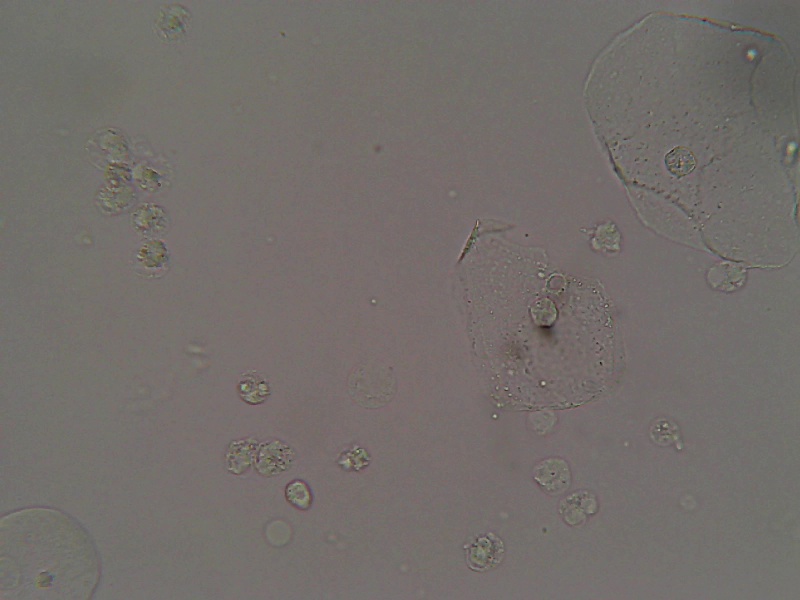}};
		\node[label={above: $13^{th}$ frame}] at (1,0) {\ig{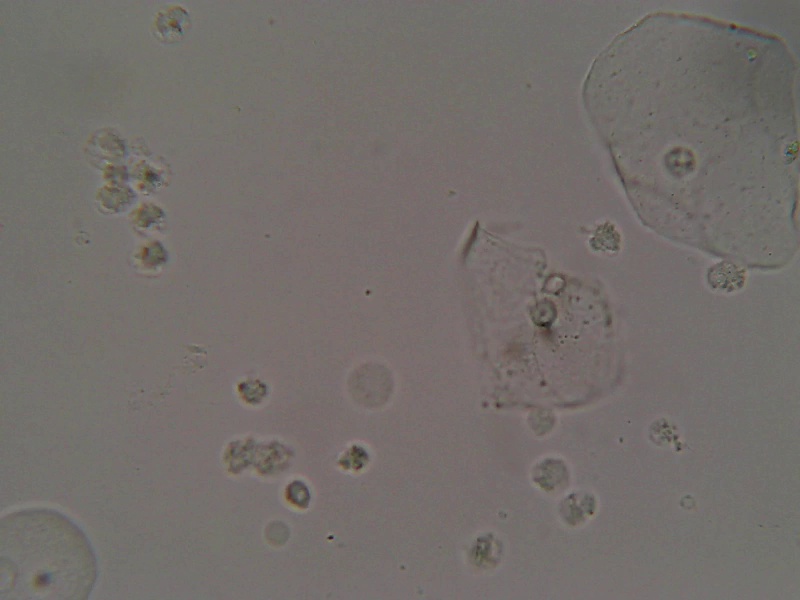}};
		\node[label={above: $20^{th}$ frame}] at (2,0) {\ig{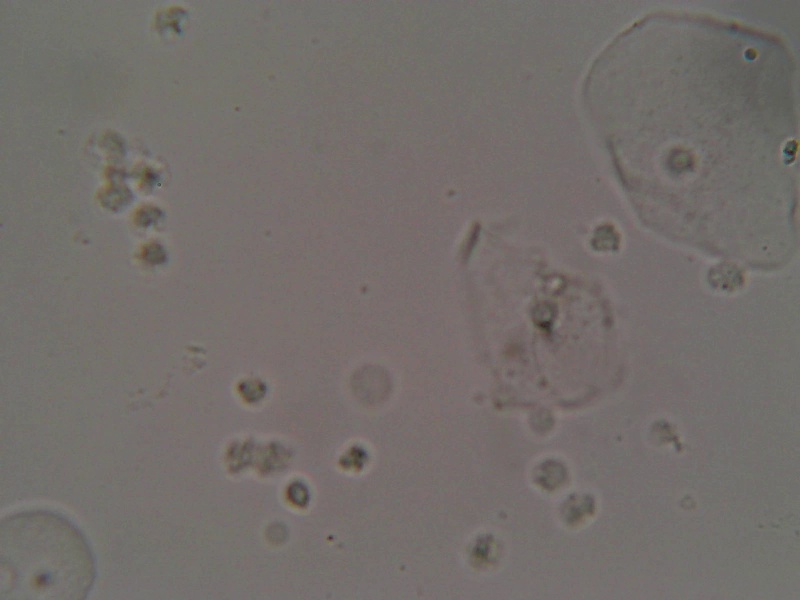}};
		\node[label={above: $34^{th}$ frame}] at (3,0) {\ig{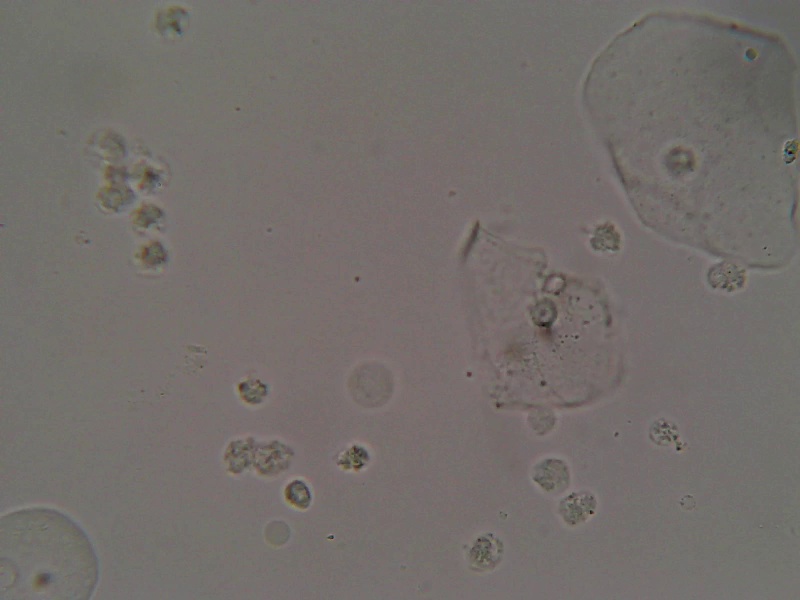}};
		\node[label={above: $43^{th}$ frame}] at (4,0) {\ig{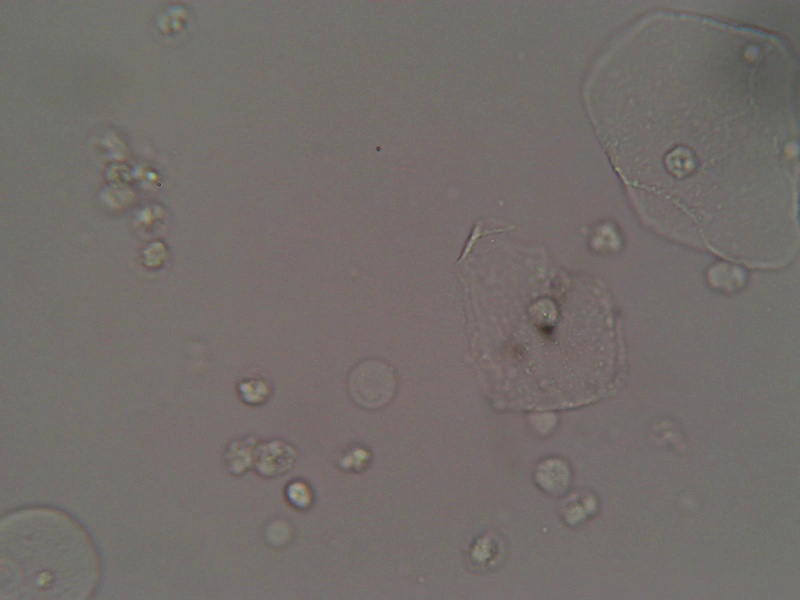}};
		\node[label={above: $55^{th}$ frame}] at (5,0) {\ig{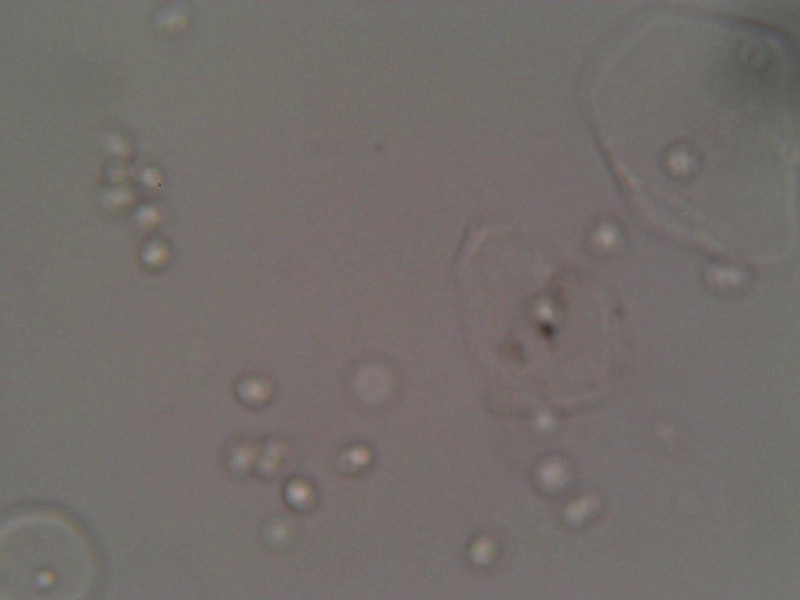}};

		\node[label={above: $0^{th}$ frame},label={[anchor=south, rotate=90]left:SL Video}] at (0,-1) {\ig{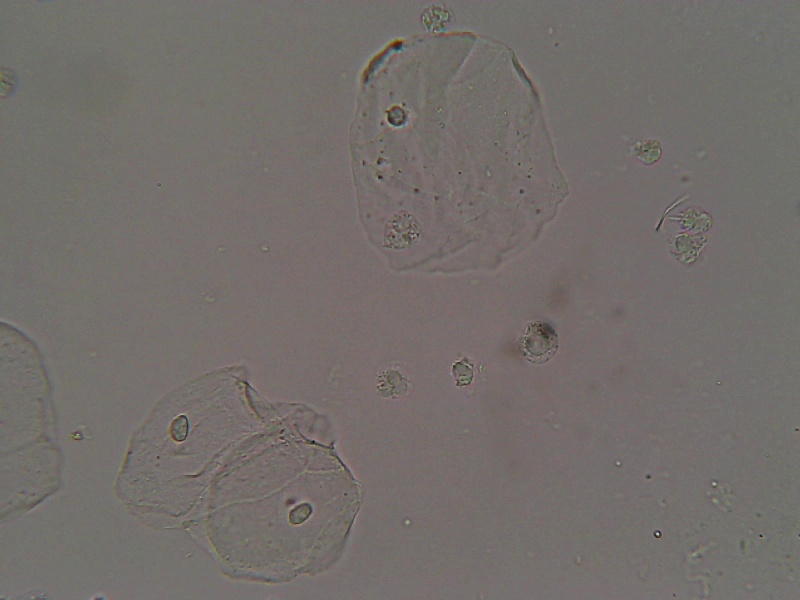}};
		\node[label={above: $11^{th}$ frame}] at (1,-1) {\ig{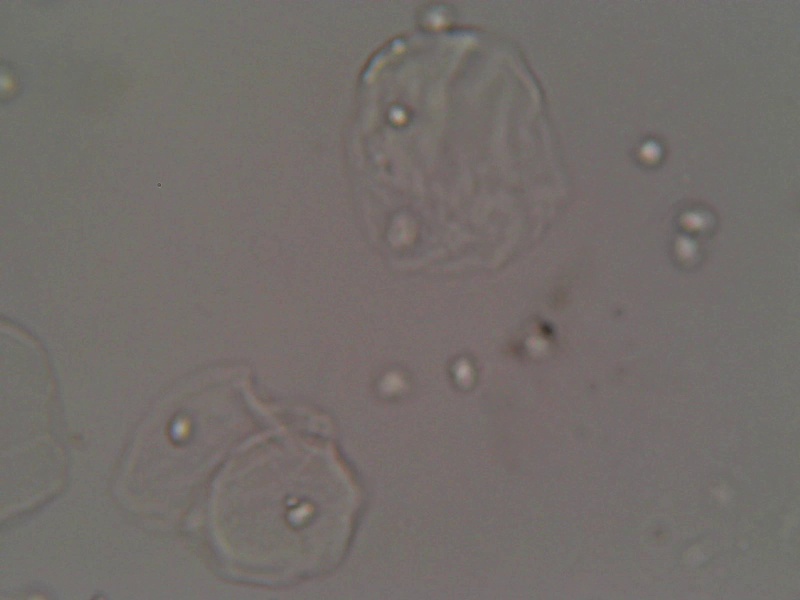}};
		\node[label={above: $25^{th}$ frame}] at (2,-1) {\ig{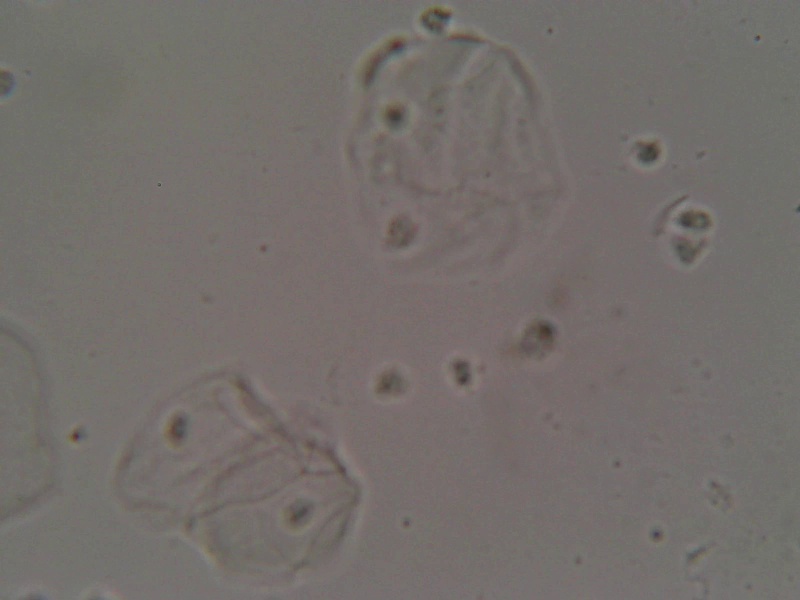}};
		\node[label={above: $35^{th}$ frame}] at (3,-1) {\ig{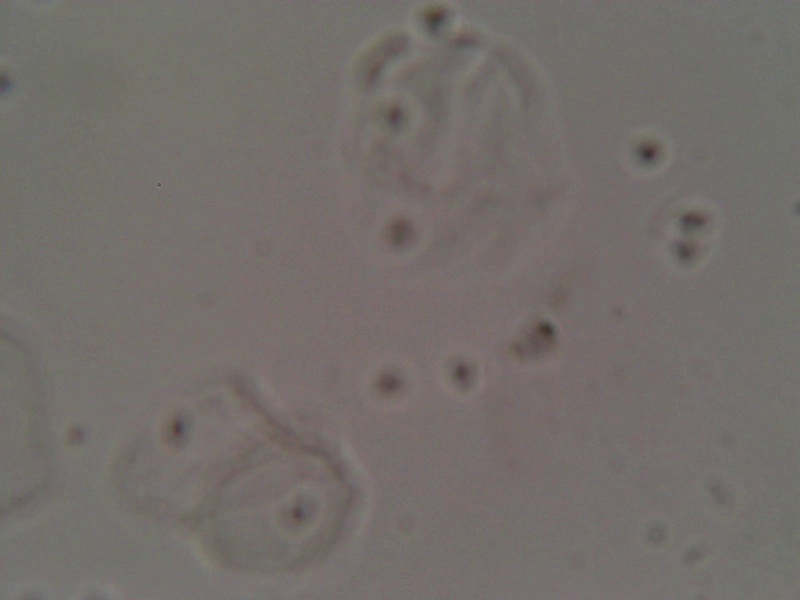}};
		\node[label={above: $49^{th}$ frame}] at (4,-1) {\ig{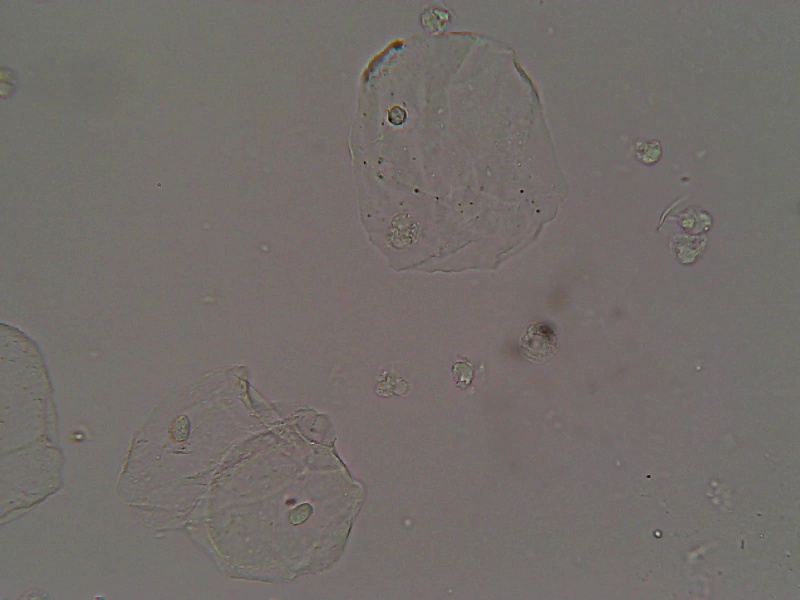}};
		\node[label={above: $53^{rd}$ frame}] at (5,-1) {\ig{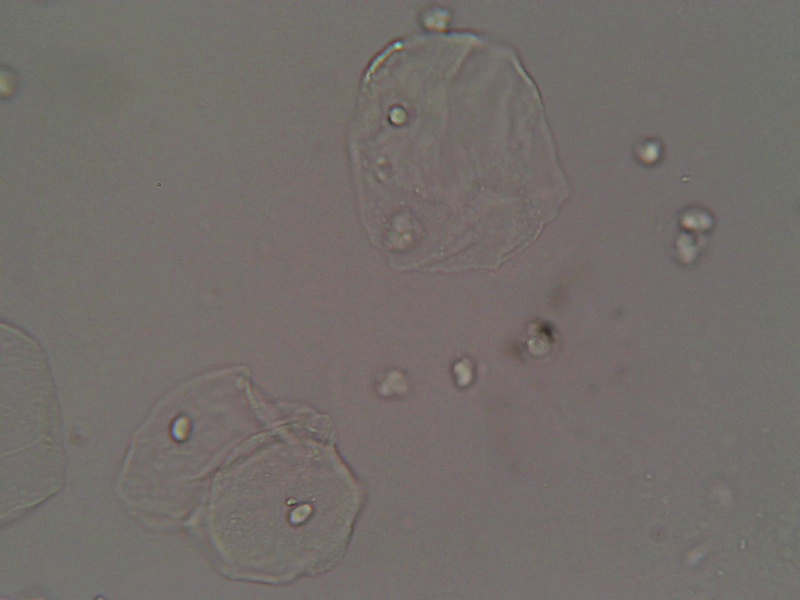}};

		\node[label={above: $0^{th}$ frame},label={[anchor=south, rotate=90]left:MV Video}] at (0,-2) {\ig{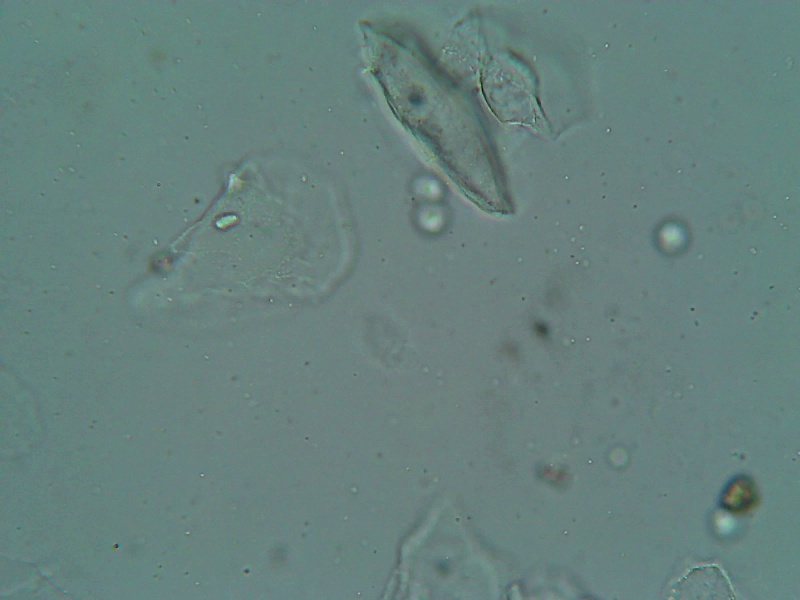}};
		\node[label={above: $49^{th}$ frame}] at (1,-2) {\ig{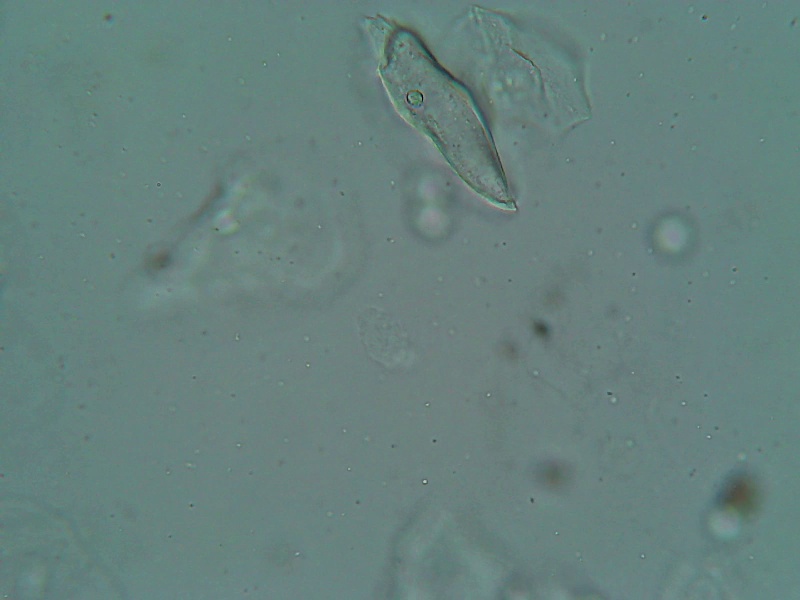}};
		\node[label={above: $123^{st}$ frame}] at (2,-2) {\ig{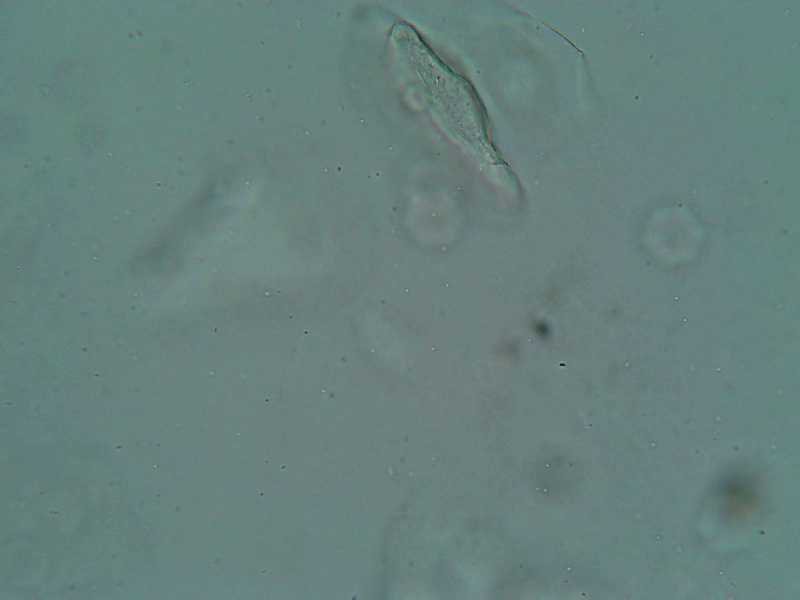}};
		\node[label={above: $170^{th}$ frame}] at (3,-2) {\ig{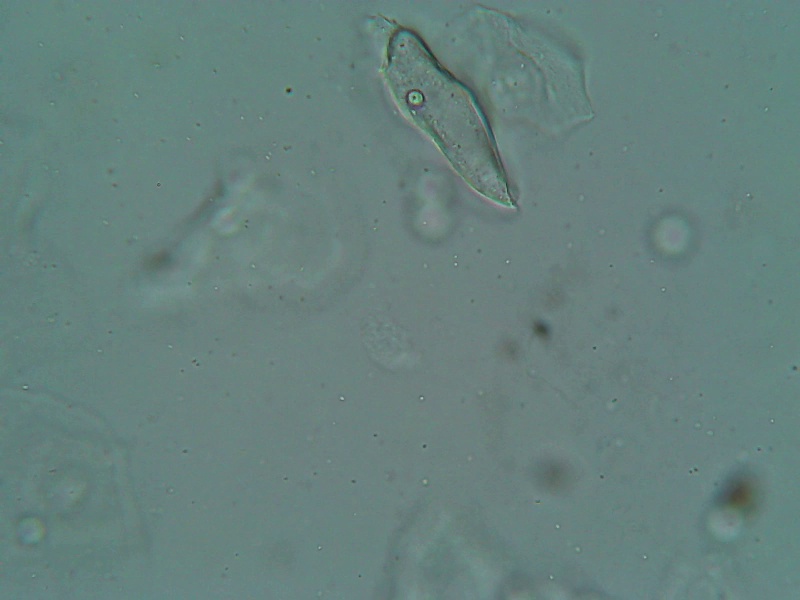}};
		\node[label={above: $241^{th}$ frame}] at (4,-2) {\ig{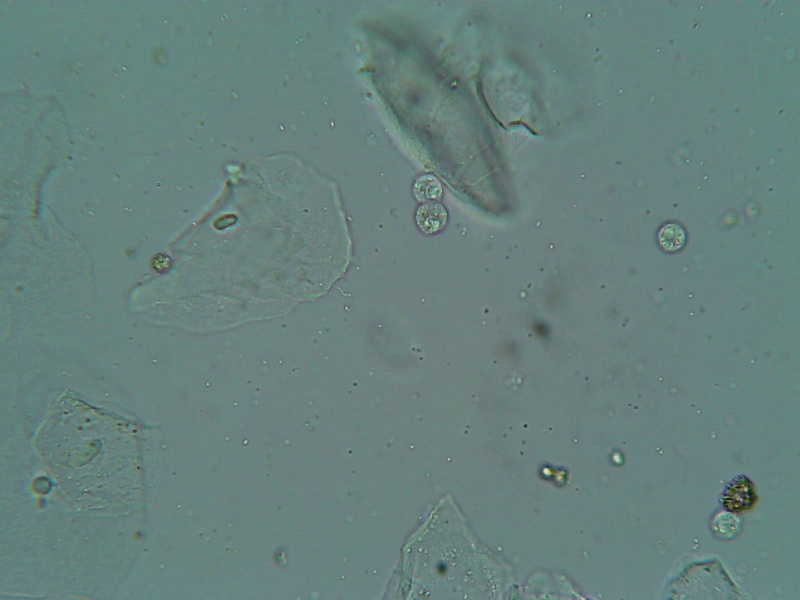}};
		\node[label={above: $276^{th}$ frame}] at (5,-2) {\ig{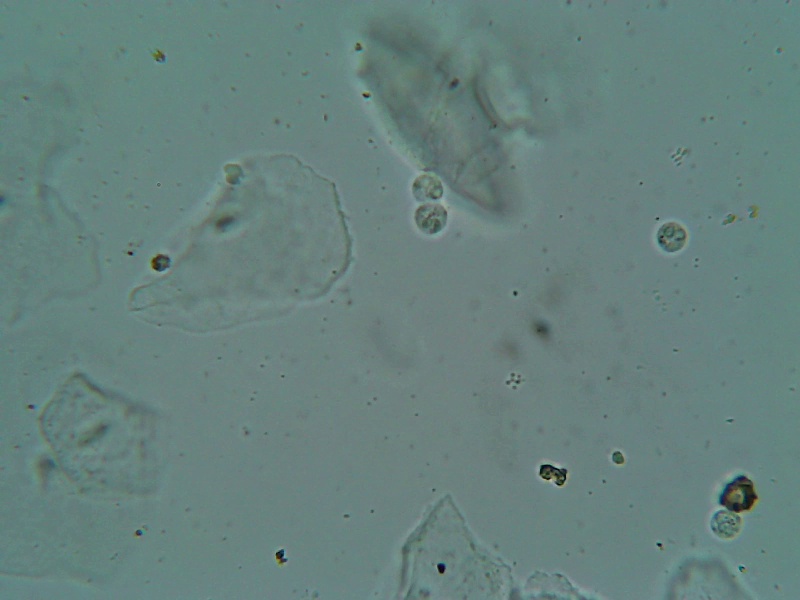}};
	\end{tikzpicture}
    \caption{DNN predictions on all-in-focus reconstructed images from 3 videos. The first row shows the ground truth annotations (green boxes with cell-class) along with predictions from DNN (red boxes with class and confidence score). The other three rows demonstrate at least one full cycle of focus variation using a few frames for each video.} 
    \label{fig:recn}
\end{figure*}

\begin{table}[t]
	\setlength{\tabcolsep}{2.5pt}
	\caption{Performance measures of predictions on all-in-focus images at confidence score $\geq 0.6$.}
	\label{tab:metric}
	\centering
		\begin{filecontents*}{data/metric.csv}
class,GT,TP,FP,FN,Prec,Recall,F1
pus,78.0,55.0,2.0,23.0,0.9649122807017544,0.7051282051282052,0.814814814814815
rbc,21.0,15.0,11.0,6.0,0.5769230769230769,0.7142857142857143,0.6382978723404256
ep,22.0,17.0,1.0,5.0,0.9444444444444444,0.7727272727272727,0.85
\end{filecontents*}
\pgfplotstabletypeset[
		col sep=comma,
        column type = c,
        columns/class/.style={string type},
        columns/Recall/.style={precision=2,zerofill},
        columns/F1/.style={precision=2,zerofill},
        every head row/.style={%
	        output empty row,
			before row={
			\toprule
			Class & Ground & True & False & False & Precision & Recall & $F_1$\\
			& Truths & Positive & Positive & Negative & & &\\
			},
			after row = {
			\midrule
			}},
		every last row/.style={after row=\bottomrule},
		]{data/metric.csv}		
\end{table}


Our dataset has mainly three types of videos: multi-layered (ML), single-layered (SL), and with moving cells (MV). In ML videos (number 0 to 7), the sharpest appearance of cells exists at different frames. In SL videos (number 8-12), more than 95\% of cells seems to appear sharp visually at a single frame of the video. The MV video number 13 is also a ML video but in this case, cells are moving slightly over time whereas in other videos, cells are almost stationary; see Fig.~\ref{fig:recn} for a few frames of ML, SL, and MV videos.

For SL videos, ideally the patch factor $p$ equal to 1 is sufficient. But, to demonstrate the robustness of our algorithm, we tried to use almost the same set of hyperparameters for all videos. The hyperparameters are selected such that the cells in the all-in-focus reconstructed image do not have any discontinuity and most of the cells are in-focus; see reconstructed images in the first row of Fig.~\ref{fig:recn}. 

The DNN predictions on reconstructed images are satisfactory except for RBC cells. The positive predictive value (precision) is close to 95\%, true positive rate (recall) is above 70\%, and $F_1$ score is around 0.81 for pus and epithelial cells; see Tab.~\ref{tab:metric}. DNN is unable to detect pus cells present in clusters which is a limitation of DNN. 

RBC predictions have high rate of false positives that leads to 58\% precision score. The DNN classifies out-of-focus small cells and epithelial nucleus as RBC due to their resemblance with smooth surface of RBC; see predictions in Fig.~\ref{fig:recn}.

The reconstruction for MV video-13 is satisfactory; but the robustness of our algorithm is uncertain for such videos.

\section{Conclusion}\label{Sec:conclusion}
Image quality is a crucial factor to build a robust and clinically meaningful automated microscopic urinalysis system. Due to multi-layered structure of a urine sample, out-of-focus cells in images are unavoidable. The proposed video based urinalysis pipeline constructs an image having most of the cells in focus with sharp image features which is essential for deep neural network models to achieve high accuracy.

The proposed patch-based all-in-focus reconstruction algorithm is robust to hyper parameters and, with DNN, our method achieves up to 95\% positive predictive value and 70\% true positive rate. Our video-based automated urinalysis system is a cheaper alternative to expensive autofocus system and useful for in-experienced technicians in villages.

We believe that the video based urinalysis could be an important area of research. In future, we plan to improve our method by tracking cells among frames and develop an open-source web application for labs to use our technology.

\bibliographystyle{ieeetr}
\bibliography{ref}

\end{document}